\documentclass[review]{elsarticle}

\usepackage{lineno,hyperref}

\journal{Journal of \LaTeX\ Templates}

\usepackage{rotating}
\usepackage{makecell}
\usepackage{mathtools}
\usepackage{xcolor}

\usepackage[linesnumbered,ruled,vlined]{algorithm2e}
\usepackage{algpseudocode}

\usepackage{comment}

\begin{document}

\begin{frontmatter}

\title{Automatically generating decision-support chatbots based on DMN models\tnoteref{mytitlenote}}
\tnotetext[mytitlenote]{
This work has been partially supported by grant PID2021-126227NB-C21 funded by MCIN/AEI /10.13039/501100011033/FEDER, EU, by grant TED2021-131023B-C22 funded by MCIN/AEI/10.13039/501100011033 and by the European Union “NextGenerationEU”/PRTR.}


\author[mymainaddress,mysecondaryaddress]{Bedilia Estrada-Torres\corref{mycorrespondingauthor}}
\cortext[mycorrespondingauthor]{Corresponding author}
\ead{iestrada@us.es}

\author[mymainaddress,mysecondaryaddress]{Adela del-R\'io-Ortega}
\author[mymainaddress,mysecondaryaddress]{Manuel Resinas}

\address[mymainaddress]{Departmento de Lenguajes y Sistemas Inform\'aticos, Universidad de Sevilla, Seville, Spain}
\address[mysecondaryaddress]{SCORE Lab, I3US Institute. Universidad de Sevilla, Seville, Spain}

\begin{abstract}

How decisions are being made is of utmost importance within organizations. The explicit representation of business logic facilitates identifying and adopting the criteria needed to make a particular decision and drives initiatives to automate repetitive decisions. The last decade has seen a surge in both the adoption of decision modeling standards such as DMN and the use of software tools such as chatbots, which seek to automate parts of the process by interacting with users to guide them in executing tasks or providing information. However, building a chatbot is not a trivial task, as it requires extensive knowledge of the business domain as well as technical knowledge for implementing the tool. 
In this paper, we build on these two requirements to propose an approach for the automatic generation of fully functional, ready-to-use decisions-support chatbots based on a DNM decision model. 
With the aim of reducing chatbots development time and to allowing non-technical users the possibility of developing chatbots specific to their domain, all necessary phases for the generation of the chatbot were implemented in the \textit{Demabot} tool. 
The evaluation was conducted with potential developers and end users. 
The results showed that Demabot generates chatbots that are correct and allow for acceptably smooth communication with the user. Furthermore, Demabots's help and customization options are considered useful and correct, while the tool can also help to reduce development time and potential errors. 


\end{abstract}


\end{frontmatter}


 \section{Introduction}
\label{sec:introduction}

Chatbots are software tools designed to interact with users through friendly conversations using natural language to simulate smooth interactions with a human \cite{Hussain_2019_SurveyChatbots,Valtolina_2108_DomainOfUse}. In recent years, the use of chatbots as a mechanism for automating business processes has increased significantly \cite{Janssen_2022_Chatbots}. Many companies are transferring their workload to them, facilitating both internal and external communications \cite{Sorin_2021_RepetitiveDecs}. In some scenarios, they can act as teammates of human workers in a collaborative way in complex processes~\cite{Seeber_2020_chatbotsMates}. In the healthcare area, for example, they have taken an active role in the provision of prevention, diagnosis, and treatment services \cite{Jovanovic_2021_ChatbotsHealth,Mujeeb_2017_Aaquabot}. Chatbots like \cite{web_chatbot_vasco,web_chatbot_riskassessment,web_chatbot_hispabot} provided great help during the emergency situation by COVID-19 as they allowed for initial screenings based on a patient's symptoms (if the patient has a fever, cough, etc.) and their contacts with infected people \cite{Brandtzaeg_2017_WhyUseChatbot}. 
Those chatbots avoid having to develop dedicated applications, making them more accessible to all kinds of users and reducing the workload in health centers. 

The benefits of chatbots have also been shown in other areas such as marketing, education, business, and e-commerce \cite{Shawar_2007_Chatbotuseful}. They have been used for a variety of purposes such as provide information, solve doubts, or help in the achievement of a specific task~\cite{Valtolina_2108_DomainOfUse}, as well as to facilitate communications \cite{Sorin_2021_RepetitiveDecs}, and in some scenarios, act as teammates of human workers in a collaborative way in complex processes~\cite{Seeber_2020_chatbotsMates}. 


Decision management is crucial for the achievement of strategic and business objectives in any organization. \textit{Strategic decisions} are decisions of great value, but they are usually made very few times in a period of time; whereas \textit{operational decisions} are often made hundreds or thousands of times during a day \cite{Fish_2012_AutomationKnowledge}. Although the latter may individually be considered of small value, when viewed as a whole they may be equal in importance to strategic decisions. 

The constant quest for process optimization has led organizations to try to automate a large part of their processes, including the decisions involved. Automating decisions seeks to improve the accuracy and consistency of decision making; it requires identifying and specifying the knowledge required to make those decisions and encoding it in an executable form \cite{Fish_2012_AutomationKnowledge}. 
In process-oriented organizations, the DMN standard~\cite{OMG_DMN_2019} is used to model and manage repetitive decisions of day-to-day business operations. DMN has been used in a wide variety of areas where processes have strong regulatory requirements such as financial services and industry, insurance, energy, technology and information services, health care, disaster management, retailers or logistics \cite{Figl_2018_DMN}. 

Although not all process activities are likely to be automated \cite{Dumas_2018_Fundamentals}, repetitive tasks have a high potential to be automated \cite{Sorin_2021_RepetitiveDecs}. Operational and repetitive decisions usually follow strict guidelines and policies, and are also characterized by requiring a limited set of data. This information usually comes from users/customers through standard mechanisms like forms \cite{Fish_2012_AutomationKnowledge}. 
In this sense, the use of artificial intelligence techniques, automated tools for the execution of tasks within business processes, and the change in communication strategies between enterprises and customers have encouraged a large number of organizations to implement virtual assistants such as chatbots that provide information, solve doubts or help in the achievement of a specific task \cite{Valtolina_2108_DomainOfUse}. 

Despite the rise in the use of chatbots, platforms for their development have mainly focused on abstracting away many details related to the natural language processing through the automated recognition of user intents, and in providing a general framework in which the conversation flow can be defined \cite{Ahmad_2020_Chatbots}. However, a chatbot developer is still needed to implement a specific conversation flow, to deal with many low-level details about parameters, to provide a set of training phrases for each conversation steps, and to provide a generic set of fallback options that guides users who do not know the possibilities the chatbot offers, amongst others. Implementing these tasks is usually time-consuming and error-prone even for simple chatbots~\cite{Gwendal_2019_Framework}.

In this article, we seek to use chatbots as a mechanism for automating decision-making, which we call ``decision-support chatbots". The usefulness of these chatbots lies in the ease with which a non-technical user can generate a chatbot from a specific domain (no-code) and stands out in processes in which the end user actively participates in making a decision and is asked for certain information. For example, when a customer indicates their preferences for the suggestion of a product or when in a health emergency situations (e.g. COVID-19) many healthcare services implemented apps that ask people about their symptoms and their contacts with infected people, like \cite{Brandtzaeg_2017_WhyUseChatbot}, providing information to end and helping to decrease the medical workload. These interactions can be easily performed by a decision-support chatbot, as shown in \cite{web_chatbot_vasco}. 

Considering that chatbots with a similar purpose share many aspects in their structure and conversation flow, there are many elements that could be reused and even their implementation could be automated. Therefore, in this article we present an approach for the automatic generation of fully functional and ready-to-use decision support chatbots, thus allowing chatbot developers to focus on the chatbot's domain features and not on low-level details related to its implementation. 
This proposal takes as input a decision model described in the DMN standard. From such a model, our approach involves a fully automatic analysis for the identification and classification of the DMN model elements and the proper values derived from the domain, their transformation as part of the chatbot components, the assembly of all chatbot components, the generation of training phrases and its integration with a natural language understanding platform (\textit{Dialogflow}\footnote{https://dialogflow.cloud.google.com}) for the analysis of user utterances, up to the generation of a web environment for user-chatbot conversation. 
In addition, the generated chatbots are used to satisfy conversational requirements that optimize the conversation with the user. For example, reducing the number of chatbot-user interactions when possible, providing help for possible user queries, and being able to recognize as much data as possible after a user utterance. All those phases of the automatic chatbot generation process were implemented in the \textit{Demabot} tool.

To validate this approach, an evaluation workshop was held with the participation of 15 people to analyze the phases of the chatbots generation process, the conversational characteristics of the chatbot, as well as the correctness and usability of the resulting chatbot.

The rest of this paper is structured as follows. Section \ref{sec:running_example} describes a running example for making a decision based on a set of fixed characteristics. 
Section \ref{sec:background} introduces concepts related to our proposal. Section \ref{sec:behavior} describes the characteristics expected in the conversational behavior of chatbots. 
Section \ref{sec:approach} presents in detail the process of automatic generation of chatbots that shapes our proposal, and Section \ref{sec:demabot} describes the implemented tool. 
Section \ref{sec:evaluation_scenario} reports on the evaluation conducted. 
Section \ref{sec:related_work} identifies related work to our proposal. 
Finally, Section \ref{sec:conclusions_futurework} concludes the article and outlines future lines of work.

 \section{Running Example}
\label{sec:running_example}

\begin{figure}[thb]
	\centering
	\includegraphics[trim={0cm 0cm 0cm 0cm},clip,width=\linewidth]{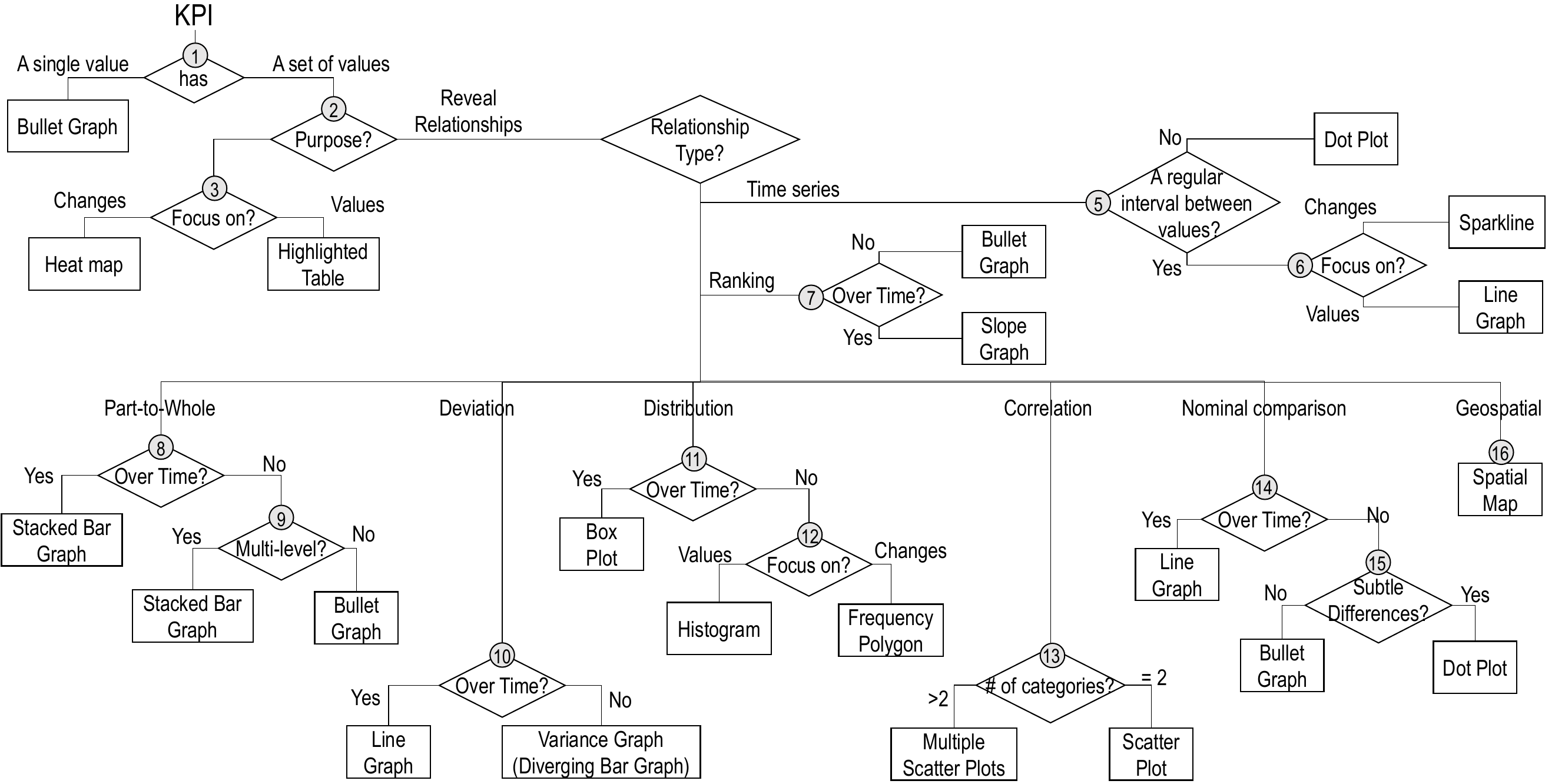}
	\caption{The decision tree used for determining the visualization of KPIs, taken from \citep{Unal_2019_KPI}}
	\label{fig:decision_tree}       
\end{figure}

In this section, we present a decision-making scenario whose goal is to determine the most appropriate type of visualization to represent the evaluation of a given key performance indicator (KPI) in a dashboard. 
The type of visualization is determined according to the KPI characteristics to be evaluated. Figure~\ref{fig:decision_tree}, taken from \citep{Unal_2019_KPI}, shows the decision tree for the 15 types of visualization analyzed: \textit{Bullet graph, Heat map, Highlighted table, Stacked bar graph, Line graph, Variance graph, Box plot, Histogram, Frequency polygon, Slope graph, Dot plot, Spark line, Multiple scatter plot, Scatter plot} and \textit{Spatial map}.  

The characteristics to take into account when making the visualization decision for the KPI are: 
(i) \textit{Single/multiple values}, whether the KPI has a single value or a set of values; (ii) \textit{Purpose}, whether the KPI is about looking up its values or revealing the relationship between its values; (iii) the \textit{Focus} of the KPI: look up changes, look up values, relationship changes in a time series, relationship values in a distribution; (iv) \textit{Relationship}, to indicate how the values of the KPI are related: time series, correlation, ranking, part-to-whole, nominal comparison, or distribution; (v) Whether the KPI needs to be displayed \textit{Over time}, (vi) \textit{Multilevel}, whether there is a hierarchy in the categories attribute of the KPI, for example, main group: continent and sub-group: country; (vii) \textit{Interval between values}, whether the KPI has regular interval between values; and (viii) whether the KPI has \textit{Subtle difference} threshold for its values.

Following the decision tree presented in \citep{Unal_2019_KPI}, for example, if we have a KPI that manages a \textit{set of values}, whose \textit{purpose} is to \textit{reveal relationships}, the \textit{relationship} between its values is \textit{time series}, the KPI has a \textit{regular} interval between its values, and its \textit{focus} is to reflect \textit{changes},  it would follow that the recommended visualization type is \textit{Spark line}. For this KPI all other characteristics are irrelevant. 

This scenario illustrates that given a set of decision rules, for example, based on the decision tree in Figure \ref{fig:decision_tree} where we specify which attributes determine the type of visualization that should be used to represent a set of data, a fully functional decision-support chatbot can be automatically built. Once that chatbot has been generated, it will be able to interact with a user indicating and requesting the information to suggest the best type of visualization depending on the data provided by the user, requesting only the strictly necessary information for decision making, in this way reducing the number of interactions with the user and providing help on the type of valid data expected by the user as answer. 
The structure generated is intended to be sufficiently generic so that in case of changing the problem domain (in this case, the suggestion of a type of visualization), the user only has to change the decision rules. To make this all work, it is necessary to generate a set of components. These will be discussed in detail in the next section.
 \section{Background and Key Concepts}
\label{sec:background}

In this section, we introduce concepts related to our proposal. First of all, we explain the main DMN elements and how they relate to our running example. Next, we describe, the key concepts related to chatbot implementations. 


\subsection{Basics of DMN}
\label{subsec:background_DMN}

The Decision Model and Notation (DMN) \cite{OMG_DMN_2019} standard emerged with the objective of providing a common notation for describing and modeling repeatable decisions in a readily understandable way for users with different roles within an organization. 
DMN deals with decision models consisting mainly of a decision table, or sets of decision tables defined in a hierarchical and interdependent manner.  A decision table is used for the definition of expressions, calculations, and if/then/else logic, among other. 

\begin{sidewaysfigure}
	\centering
	\includegraphics[trim={0cm 0cm 0cm 0cm},clip,width=\linewidth]{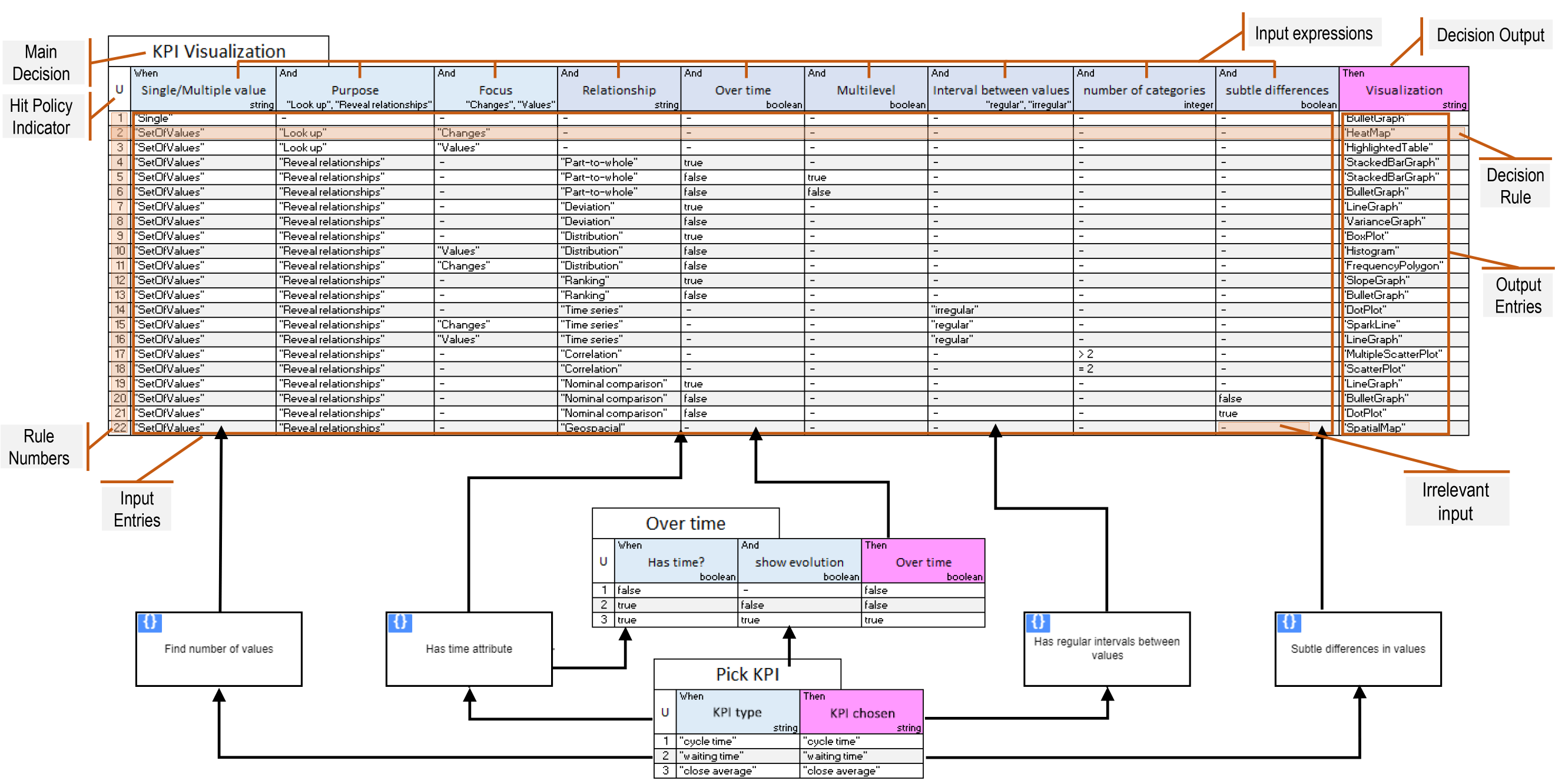}
	\caption{Example of a DMN model - Hierarchy of decision tables based on the decision tree proposed in \citep{Unal_2019_KPI}}
	\label{fig:decision_model}       
\end{sidewaysfigure}

To illustrate the main DMN elements that we will focus on in this article and to facilitate the understanding of our proposal, we took as a basis our running example to build the DMN model shown in Fig.~\ref{fig:decision_model}. 
This DMN model is a hierarchical structure consisting of three \textit{decision tables} (\textit{KPI Visualization}, \textit{Over time}, and  \textit{Pick KPI}) and four \textit{literal expressions} \textit{(Find number of values}, \textit{Has time attribute}, \textit{Has regular intervals between values}, and \textit{Subtle differences in values}). We used this structure to show the possibility of using a hierarchy of table and also to simplify the model, since using a single table would require a large number of decision rules with many repeated attribute values that would make the example difficult to understand.

A \textit{decision table} defines the possible combinations of the different criteria required to determine a decision. 
More specifically, the main \textit{decision} \textit{KPI visualization} is resolved using a set of \textit{decision rules} (rows from 1 to 22 in Fig.~\ref{fig:decision_model}). 
Each \textit{decision rule} of this table is composed of a set of \textit{input} (9) and \textit{output entries} (1) and is identified by a \textit{rule number}. The \textit{hit policy indicator} indicates what the result of the decision table is in cases of rule overlapping, that is, when more than one rule matches the input data. By default, and in our example, hit policy \textit{U, unique} is used, in which case it is assumed that the table has no overlaps. 

Decision tables can be defined horizontally or vertically. In our example, they are all vertical, where each column with \textit{input expression} (Single/Multiple value, Purpose, Focus, etc.) represents a type of \textit{input entry}; and the last column (Visualization) represents the \textit{decision output}. 
All input entries in \textit{Pick KPI} decision are required to make the decision, but in the other two tables there are entries whose value is not relevant for the decision making, called wildcards (``-"). For example, the \textit{show evolution} entry in rule 1 for decision \textit{Over time} or the \textit{Relationship} entry in rules 1-3 for decision \textit{KPI Visualization}.  

The \textit{literal expressions} represent decision logic as a text that describes how an output value is derived from its input values using literal statements or function invocation. Table~\ref{tab:literal_expression_values} shows the values provided by each literal expression, depending on the given type of KPI, which can be one of three values: \textit{cycle time}, \textit{waiting time}, and \textit{close average}.

\begin{table}[htb]
    \scriptsize
    \centering
    \caption{Values provided for each decision literal expression depending on the given KPI}
    \begin{tabular}{c c c c c} \hline
      \textbf{KPI}  &  \makecell{\textbf{Find number}\\ \textbf{of values}} &  \makecell{\textbf{Has time}\\ \textbf{attribute}}  &  \makecell{\textbf{Has regular intervals}\\ \textbf{between values}} & \makecell{\textbf{Subtle differences}\\ \textbf{in values}} \\ \hline
        cycle time      & 1     & false & false & 0 \\ 
        waiting time    & 12    & true  & true  & 0.3 \\
        close average   & 12    & false & false & 0.09 \\ \hline
    \end{tabular}
    \label{tab:literal_expression_values}
\end{table}


\subsection{Basics of Chatbots}
\label{subsec:background_Chatbots}

\begin{figure}[thb]
	\centering
	\includegraphics[trim={0cm 0cm 0cm 0cm},clip,width=\linewidth]{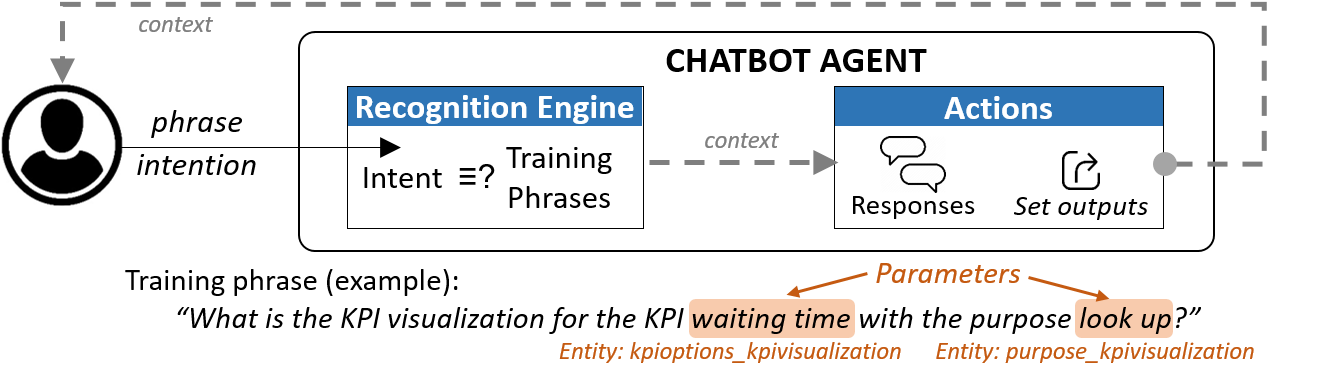}
	\caption{Interaction between key concepts of chatbots}
	\label{fig:key_concepts}       
\end{figure}

Chatbot design typically relies on parsing techniques, pattern matching strategies and Natural Language Understanding (NLU) to process user inputs. The latter has become the dominant technique thanks to the popularization of libraries and cloud-based services such as \textit{Dialogflow}, \textit{wit.ai}\footnote{https://wit.ai/} or LUIS\footnote{https://www.luis.ai/}, which rely on Machine Learning and Natural Language Processing techniques to understand user input \citep{Gwendal_2019_Framework}. Figure~\ref{fig:key_concepts} represents interactions between the key concepts of chatbots described below.

According to \citep{Gwendal_2019_Framework}, an NLU chatbot, also known as \textit{agent} \citep{Valtolina_2108_DomainOfUse}, contains a recognition engine that matches user \textit{inputs} with \textit{intentions (intents)} during a conversation turn. It also contains an execution component capable of executing \textit{actions} for each intent, such as \textit{responses} to the user. 

\textit{Intentions} or \textit{Intents} are defined through \textit{training phrases}, which are input examples that allow the recognition engine to identify the different phrases that a user can utilize to express an intention. 
For instance, an intent can be defined to recognize that the user wants to make a decision. Some training phrases for such an intent could be ``I want to determine the KPI visualization,'' or ``What is the KPI visualization for the KPI close average with changes and distribution?'' In each \textit{training phrase}, concrete values can be recognized. These values are called \textit{parameters}. For instance, in the second example, there are three parameters, namely: KPI  equals to `close average', purpose equals to `look up' and relationship equals to `distribution'. The type of the parameters is defined by a specific structure called \textit{entity}, which determine how data from a user input is extracted. NLU platforms provide predefined entities that match many common types of data like numbers, dates, times, colors, or e-mail addresses. In addition, it is possible to define custom entities for enumerated values. A custom entity is composed of a set of entries. Each entry is made up of a reference value and a set of synonyms for that reference. For example, if we define a ``boolean'' entity, two entries are required: \textit{true} and \textit{false}. For the \textit{true entry}, the values \textit{yes, ok, correct}, could be synonyms for it. 

Each intent can be associated to \textit{input} and \textit{output contexts} to control the flow of the conversation. 
In addition, contexts store the parameter values provided by a user in one intent so that it can be used in other intents.
Finally, for each intent that is recognized one action is executed: typically sending a response to the user and or setting values to make new utterance.

 \section{Characteristics of the conversational behavior of chatbots}
\label{sec:behavior}

The main objective of our proposal is to provide a mechanism for the automatic generation of chatbots based on a decision model defined in DMN. By using this approach, we also seek to improve the conversational characteristics of the generated chatbot. 

According to \citep{Chavez_2021_Challenges}, there are three characteristics associated with conversational behavior that should be present in chatbots: proactivity, conscientiousness and communicability. For each characteristic, benefits and challenges are proposed. \textit{Proactivity} seeks to reduce the amount of human effort to complete a task and seeks to provide the user with additional and useful information for the continuity of the conversation, as well as to redirect the conversation in case of error and reduce the time in obtaining a response. In addition, it is recommended to inform the user of the context of the conversation.  
With respect to \textit{conscientiousness}, it is sought that the user has a sense of continuity in the conversation despite possible errors that may arise during the conversation, the chatbot must be able to understand the purpose of the information provided by the user and strive to lead the conversation towards the goal efficiently and productively. In addition, it must manage decision complexity by minimizing interaction turns and provide information about the flow of the conversation and the steps taken. 
Finally, with respect to \textit{communicability}, the chatbot must be able to indicate the purpose of the conversation, as well as provide helpful information that facilitates communication with the user.

Based on this information and the experience of the authors of this article, we specify below a set of requirements that the generated chatbot must address. 


Requirements R1, R2 and R3 are related to \textit{conscientiousness}, R4 and R5 to \textit{proactivity}, and R6 and R7 to \textit{communicability}. 

\begin{itemize}
    \item \textbf{R1:} \textit{The information required to make the decision can be provided in any order.} In other words, the user does not have to learn a predefined structure that has to be used to provide the information to the chatbot. 
    
    \item \textbf{R2:} \textit{The user can provide several input values simultaneously}, even in the first interaction, to improve efficiency, especially for advanced users. This means that, for our example, the chatbot must be able to deal with phrases like ``What is the \textit{KPI visualization} (decision) for the KPI type \textit{waiting time} \texttt{(type:parameter)} with purpose \textit{look up} \texttt{(purpose:parameter)}?'' By doing so, the chatbot does not have to ask again about about the type and purpose of the KPI, lowering the number of interactions with the user and making the conversation more human-like.
    
    \item \textbf{R3:}\textit{ Only information that is necessary should be asked to the user}. This means that in a chatbot developed for the decision table in Figure \ref{fig:decision_model}, if the user said ``the KPI represents \textit{SetOfValues}, it has \textit{Look up} as purpose, and the focus is \textit{changes}'', other values such as \textit{relationship} or \textit{multilevel} should not be asked because those values are irrelevant for the decision.
    
    \item \textbf{R4:} \textit{Guide the conversation}. Redirect the conversation in case of error or misunderstanding of the user's information, as well as to provide response alternatives and/or expected response characteristics (text, number, etc.) for each chatbot question. This way the user will have an automatic help that benefits the flow of the conversation. 
    
    \item \textbf{R5:} \textit{Provide context information}. Let the user know what their previous answers have been in the context of that conversation to make a decision.
    
    \item \textbf{R6:} Whenever the user requests it, the chatbot should \textit{be able to provide} generic \textit{help} about the conversation and specific help for each of the values expected during the conversation.
    
    \item \textbf{R7:} \textit{Suggestion of valid answers. } Suggest to the user valid answer values for each question. In case the possible answer to a chatbot question is a boolean value, a number or a restricted set of words, show these valid options to the user before receiving an answer or requesting help. 
    
    
\end{itemize}
 \section{Automatic Chatbot Generation Process} 
\label{sec:approach}

\begin{figure}[thb]
	\centering
	\includegraphics[trim={0cm 0cm 0cm 0cm},clip,width=\linewidth]{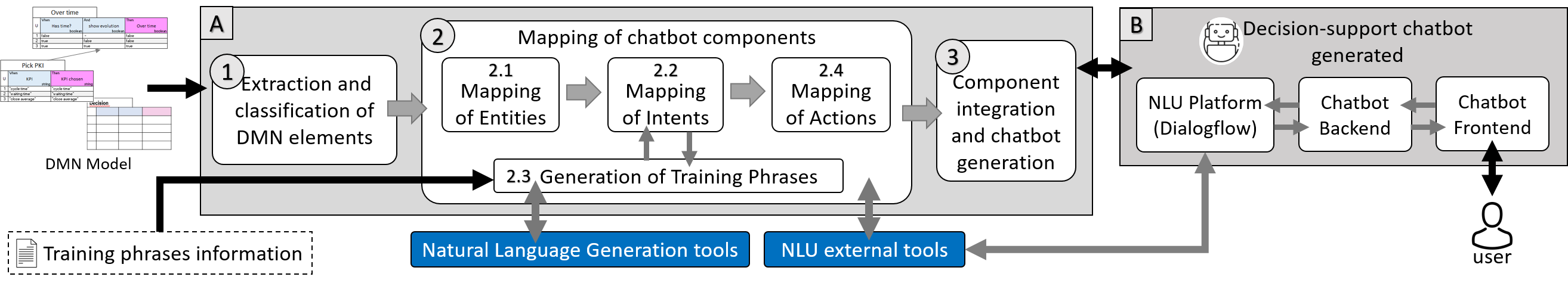}
	\caption{Phases of the automatic chatbots generation process and the generated chatbot.}
	\label{fig:process_phases}       
\end{figure}

In this section, we address the process of automatic generation of chatbots from a DMN model. Figure \ref{fig:process_phases} depicts an overview of our proposal. Block \textit{A} represents the process of identifying, analyzing and transforming DMN elements into chatbot elements. Block \textit{B} represents the components of the generated chatbot, which will be discussed in more detail in Section \ref{sec:demabot}.  The main input of the process is a DMN decision model. Optionally, information for the generation of training phrases can also be provided. Phase \textit{1} is responsible for analyzing this model to extract and classify the DMN elements and domain specific values. With these components, phase \textit{2} performs a mapping between the DMN elements and the components necessary for the implementation of the chatbot (entities, intents, training phrases and actions), interacting with external tools both for the generation of training phrases and for the storage of the chatbot components. Phase \textit{3} consists of the integration of all these components and the generation of an interface that allows the interaction with the generated bot agent. As a result we have a fully functional decision-support chatbots. 
The generated chatbot (block \textit{B}), interacts with the NLU platform to store the chatbot structure and obtain context information, the background is the engine in charge of determining the conversation flow and performing domain validations. It also interacts with the frontend (interface), which is the one that is presented to the user to carry out the conversation that provides the solution to a decision. 

Below we describe each of these phases in more detail.
Below we describe each of the phases of the automatic chatbot generation process (block A) in more detail. 

\subsection{Extraction and Classification of DMN Elements}
\label{subsec:approach_extractionClassificationElements}

From the DMN model it is necessary to identify all the decision tables and literal expressions defined in it. Afterwards, it is necessary to determine which is the main decision (KPI visualization, for the example of Fig.~\ref{fig:decision_model}). For each decision table, it is necessary to identify the dependencies. For example, the decision table \textit{Pick KPI} has no dependencies, that means that to obtain an output all the values depend directly on the values of its inputs. However, the decision \textit{Over time} requires the output of the decision table \textit{Pick PKI} and the result of the literal expression \textit{Has time attribute}. 
For each input expression it is necessary to identify the expected data type and the set of all its possible values. For example, the \textit{Purpose} attribute of the \textit{KPI Visualization} decision table is of type \textit{string} and its possible values are \textit{Look up} and \textit{Reveal relationships}, while the \textit{number of categories} input expression is of type \textit{integer}. 

\subsection{Mapping of Chatbot Components}
\label{subsec:approach_mappingComponents}

In this phase, the information obtained in the previous phase is taken as input to generate the main components to generate a chatbot: entities, intentions, parameters, contexts, actions, and training phrases. We generate this mapping of components through four sub-phases described below. 


\subsubsection{Mapping of Entities}
\label{subsubsec:approach_mapping_entities}

Each \textit{input expression} of a decision table is associated with an \textit{entity} that defines the data type associated with it. 
NLU platforms usually provide predefined entities that match common data types. For example, \texttt{number} represents ordinal or cardinal numbers, or \texttt{email} that specifies that a value complies with an email format. However, there are entities that depend on the domain being addressed or that, due to their structure, are not covered generically in NLU platforms. 

For numeric values (integer, long, double) there are usually predefined entities that allow a direct mapping, but for values such as boolean or string, it is usually necessary to define a more domain-specific $custom$ entity type. This latter mapping should take the following into account.

\begin{itemize}

\item Name the custom entity as \texttt{ent\_<inputexpression>\_<dtname>}, where \texttt{inputexpression} is the input expression being mapped and \texttt{dtname} is the name of the decision table to which the input expression belongs. 

\item If the input expression type is \textit{boolean}, the entity has two \textit{entries}: \emph{true} and \emph{false}. In addition, to make the conversation more smooth and fluid, several synonyms for them must be defined. These synonyms include the typical ``ok'' and ``yes'' for a \textit{true} value, and their false counterparts. 
Since a \textit{boolean} entity could be used in several decision tables with the same entries and synonyms, it can be defined generically as \texttt{ent\_boolean}
As an additional feature, the input name itself could also be considered as true and its negation in several forms. For instance, for the input expression \textit{Multilevel} (Figure \ref{fig:decision_model}), the entity \texttt{ent\_multilevel\_kpivisualization} must be created with two entries: \emph{True, False}, where the synonyms of \textit{True} could be \textit{yes, multilevel, has multilevel} and the synonyms of \textit{False} could be \textit{no, not multilevel, without multilevel}.

\item If the input expression type is \textit{string}, the non-repeated input entries can be considered as an \textit{enumerated} structure, where the entity has one entry for each of the enumerated values. For instance, for the input expression \texttt{KPI type} in the \texttt{Pick KPI} decision table, a decision intent \texttt{ent\_kpitype} is created. Its allowed values are the strings \textit{cycle time}, \textit{waiting time}, and \textit{close average}. For strings, enumerated values, no synonyms are provided by default, but the chatbot developer can provide them if needed.

\end{itemize}


\subsubsection{Mapping of Intents}
\label{subsubsec:approach_mapping_intents}

According to their function, and with the intention of fulfilling the identified requirements, we propose to classify intents into three types: \textit{decision intents}, \textit{input intents} and \textit{support intents}. Each one is described below.


A \textbf{Decision Intent: } is created for each decision table in the DMN model. The proposed format for naming decision intents is \texttt{<dtname>}. 
Their goal is to identify the decision that the user wants to make, for which the intent must recognize phrases that convey this intention like \textit{``I want to determine the KPI Visualization,''} or \textit{``I want to know the KPI Visualization,''} or maybe directly \textit{``KPI visualization.''}  

The decision intent must also be able to gather as many information from that user input (R2) dealing with phrases like \textit{``What is the KPI Visualization with purpose look up and focus in values.''} As a consequence, one \textit{parameter} for each input expression of the decision table is added to the intent. We propose to name each parameter by the name of the attribute it represents (\texttt{<inputexpression>}). 
The type of each parameter is either a system entity related to the type of the input or the custom entity already defined for the input. 
For instance, for the decision table \textit{Pick KPI}, a \textit{decision intent} \texttt{pickkpi} is created and it would contain one parameter (\texttt{kpitype}) of type \texttt{ent\_kpitype}. This type is related to the entity \texttt{ent\_kpitype} defined in the previous step. However, for the decision table \textit{KPI Visualization}, a \textit{decision intent} \texttt{kpivisualization} is created and it would contain seven parameters (\texttt{kpitype, showevolution, purpose, focus, relationship, multilevel, numberofcategories}), one for each input expression, excluding the those obtained from literal expressions (\texttt{singlemultiplevalue, overtime, hastime}).  

Finally, \textit{decisions} have no input context and have a predefined output context representing the chosen decision. This output context is named following the format \texttt{<dtname>\_decision}. However, other output contexts are added during the action phase, as described in the next section. 


An \textbf{Input Intents: } is created for each \textit{input expression} in the DMN model. 
The \texttt{<dtname>\_<inputexpression>} format is proposed for naming them. 
Their main goal is to gather information about each input after a question made by the chatbot (e.g., \textit{``What is the KPI?''}).
However, like in a decision intent, it is necessary to provide a mechanism to capture information about other inputs that the user might include together with the response. Therefore, the phrases that need to be recognized are like \textit{``cycle time''}, but also more complex sentences involving values from one or more decision tables such as \textit{``the kpi type cycle time with focus values and relationship as distribution.''} As a consequence, again one parameter (named as \texttt{<inputexpression>}) for each input of the decision table is created.

In this type of intent, the parameter that corresponds to the intent associated with the same input will be defined as $required$. If a \textit{required parameter} is not provided, the intent cannot continue with the normal conversation flow and will request this parameter from the user. For example, in the intent \texttt{kpivisualization\_focus}, seven parameters must be included (the same as for the intent \texttt{kpivisualization}), but only the parameter \texttt{focus} will be marked as \textit{required} and the other six will be \textit{optional} parameters.  

Input intents have two \textit{input context}: one related to the decision table associated with the input, named as \texttt{<dtname>\_decision} and the other one related to the input itself \texttt{<dtname>\_<inputexpression>}, for example, for the intent \texttt{kpivisualization\_focus}, the contexts \texttt{kpivisualization\_decision} and \texttt{kpivisualization\_focus} are created. One \textit{output context} is created for the intent, \texttt{<dtname>\_decision}. All these contexts are activated to signal the moment in which that intent is requested within the conversation flow (R5). 


The goal of \textbf{Support Intents} is to make communication with the user more fluid and friendly and provide help if necessary. In this sense, we propose two types of support intents: domain-independent and domain-dependent. 

\textit{Domain-independent support intents} are intents that can be used with the same data and structure in any chatbot, such as the following (R4, R6):
\begin{itemize}
\item \texttt{WellcomeIntent} identifies the first interaction by the user to start a conversation, such as a greeting. 
\item \texttt{FallbackIntent} is activated when the value provided by the user is not an expected value for the previous question, so that it suggests the user to provide a new value or to ask for help. 
\item \texttt{HelpIntent} is triggered when the user ask for help in a generic way, because the user does not know how to continue the conversation. 
\item \texttt{CancelIntent} identifies when the user wants to end a conversation.
\item \texttt{EndIntent} identifies when the user wants to end a conversation politely. 
\end{itemize}

\textit{Domain-dependent support intents} are intents used to identify and provide help to the user about the expected values for a particular input (R7). A \textit{domain-dependent support intents} is created for each \textit{input intent} previously created. The suggested name for these intents is \texttt{<dtname>\_<inputexpression>\_help}. This intent will have the same contexts as the intent \texttt{<dtname>\_<inputexpression>}. For example, for the intent 
\texttt{kpivisualization\_focus}, the \texttt{kpivisualization\_focus\_help} intent is created.


\subsubsection{Generation of Training Phrases}
\label{subsubsec:approach_mapping_trainingPhrases}

Each intent should contain a set of training phrases, which are input examples used to detect which intent the user refers to. These training phrases also detail how to extract the information to fill the parameters of the intent from the user's message. For instance, the sentence \textit{``I want to determine the KPI visualization for a KPI type cycle time that shows evolution  with purpose reveal relationship''} should match with the main decision intent and has to fill three parameters, namely \texttt{kpi type}, which evaluates to \emph{cycle time}; \texttt{show evolution}, which evaluates to \emph{true}, and \texttt{purpose}, which evaluates to \emph{relationship}. 

The generation of training phrases is one of the most relevant, but also more time-consuming tasks while developing a chatbot based on NLU. In our approach we provide an automatic way to generate these phrases based on natural language generation (NLG) in such a way that the information can be captured in any order provided by the user (R1). Traditionally, NLG has been used in chatbots to generate its responses to the users. However, with the advent of NLU chatbots, new NLG tools have been particularly designed for building training phrases for chatbots. The reason is that this task is slightly different from other NLG tasks because the goal is not to build a set of phrases intended for humans, but instead to build a dataset for training the chatbot. Therefore, it is not strictly necessary that the resulting phrases are fully syntactically correct because the NLU layer already deals with this aspect. Furthermore, the natural language generation must be designed to provide a wide variety of examples. There are several tools that can be used for this task. The approach followed by them is to build a generation specification that define the patterns of text that are used to create the dataset. In this paper, we use the open source project \textit{Chatito}\footnote{\url{https://rodrigopivi.github.io/Chatito/}}.

\begin{figure}[thb]
    \centering
	\includegraphics[trim={0cm 0cm 0cm 0cm},clip,width=\linewidth]{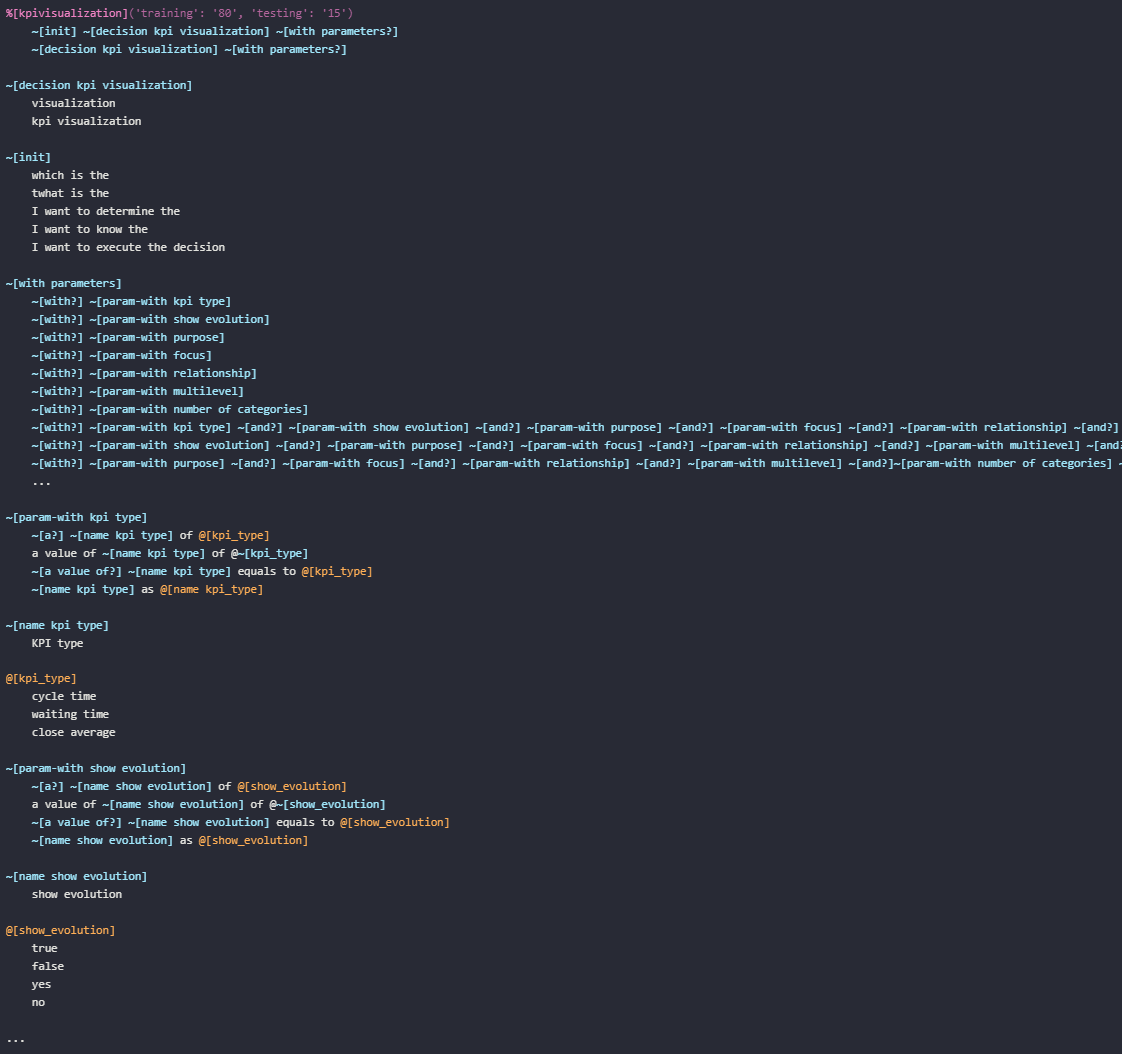}
	\caption{Natural language generation specification.}
	\label{fig:chatito_text}       
\end{figure}

Figure \ref{fig:chatito_text} depicts an extract of such a specification for the example chatbot based on the Fig.~\ref{fig:decision_model}. 
It includes an intent entity called \texttt{kpivisualization} that can be generated from two alternative sentences. 
Each of the sentences refers to alias entities like \texttt{init} or \texttt{decision} that provide alternatives for the phrase generation. Alias entities can be made optional by adding a question mark at the end of the alias name like in \texttt{\ $\sim$[with parameters?]}. Finally, slot entities, which represent parameter values, are annotated with \texttt{@}. 
The procedure to generate training phrases requires the DMN model together with some hints about how to deal with some input parameters are processed and used to automatically build a specification for each of the intents of the chatbot. Finally, the NLG tool is used to generate the set of training phrases that will be used to train each intent in the NLU chatbot. More details on the generation of these phrases are described below. 


\textbf{Generation for Decision Intents.} The pattern for decision intents is represented by an optional, initial (\texttt{[init]} phrase like ``I want to know the,''; the name of the decision (\texttt{[decision kip visualization]}), which can be either the output or the name of the decision in the DMN model, and an additional patterns to include parameters, namely: \emph{[with parameters?]}, although more patterns could be easily added following a similar approach. 
Regardless of the type of pattern, a slot entity is created for each input. The values of the slot entity depends on the type of the input and include the domain of the input or a subset if the domain is infinite like a number or a string. The only exception is if the type of the input is a boolean for which in addition to the values of ``true'' and ``false'', the synonyms of ``yes'' and ``no'' are included.  

Having the slot entities defined, the pattern of \emph{[with parameters?]} include the optional preposition \emph{with} and k-permutations of the slot entities created for each input with $k=\{1,\ldots,n\}$, where $n$ is the number of inputs.  
If the number of inputs is greater than three, then only all 1-permutations, one n-permutation and a random subset of the other k-permutations are chosen. This can be done because we are only creating training examples, not defining the whole set of possible phrases the chatbot has to recognize. 
The permutations used to generate the training phrases for the parameters allows the user to provide the information about the different inputs in any order following requirement R1. 


\textbf{Generation for Input Intents.} The pattern for input intents is composed by an alias entity that represents the answer to the question and, optionally, some additional parameters. The former includes options like providing directly the slot entity, or surrounding it with some additional text like ``\texttt{$\sim$[a value of?] $\sim$[name kpi type?] equals to @[kpi type]}''. 
For instance, after the question ``what is the kpi type?'', a possible answer could be: ``a value of KPI type equals to cycle type''

\subsubsection{Mapping of Actions}
\label{subsubsec:approach_mapping_actions}

Each intent has an associated action that is performed when the intent is recognized. 
The most straightforward action is related to \textit{support intents}, which are canned responses provided by the chatbot. For example, a \textit{domain-dependent support intents} such as \textit{EndIntent}, may answer \textit{You're welcome}, \textit{Come back soon}, etc. However, the action that needs to be implemented for \textit{domain-dependent support intents} is more elaborated. 
Our proposal uses the Algorithm \ref{alg:required_inputs}, which depicts the response action, every time a question turn occurs in which the user answers. 
It assumes we have three functions available for the decision at hand: \texttt{inputs(decision\_name)}, which returns all the inputs to make the decision; \texttt{calculate\_decision(user\_provided\_parameters)}, which returns the decision for the given set of parameters, where \texttt{user\_provided\_parameters} is a map that assigns a value to each input, and \texttt{is\_necessary(input, parameters)}, which returns whether the given input is necessary given the current values assigned to the parameters. For instance, in the example of Figure \ref{fig:decision_model}, \texttt{is\_necessary(Over time, show evolution, has time = false)} would evaluate to false because the value of \emph{show evolution} does not affect the decision, whereas \texttt{is\_necessary(Over time, show evolution, has time = true)} would evaluate to true. The execution of the functions is carried out in an external service.

Following Algorithm \ref{alg:required_inputs}, it receives as input the parameters recognized in the intent and iterates over the inputs of the decision (line 2). Then, it is checked if a parameter value has already been provided for each input (line 3). If the input is missing, it is checked whether with the current parameters, the missing input is really necessary by means of function \texttt{is\_necessary} (line 4). Therefore, only required information is asked to the user as imposed by requirement R3. If this is the case, an output context is created and activated with the name of the missing input and a response like ``What is the \textit{$<$input$>$} value?'' is sent requesting the missing value. This is performed by function \texttt{ask\_for\_missing\_parameter} (lines 5-8).
If the parameters have been provided for all the required inputs, the decision is made and a response is sent including the decision result (lines~9-11).




\SetKwInput{KwInput}{Input}                
\SetKwInput{KwOutput}{Output}              

\begin{algorithm}
\DontPrintSemicolon
\footnotesize
  \KwInput{$decision\_name, user\_provided\_parameters$}
  
  missing\_input = false
  
  \For{input in inputs(decision\_name)}{
    
    \If{input not in user\_provided\_parameters}{
        
        \If{is\_necessary(decision\_name, input, user\_provided\_parameters)}{
        
            missing\_input = true
        
            add `awaiting\_'+input as output context 
            
            send response ask\_for\_missing \_parameter(input)
            
            break
        }
    }
    }
    \If{not missing\_input}{
        decision = calculate\_decision(user\_provided\_parameters)
        
        send response `The result is: '+decision 
  
  }

\caption{Actions of decision and input intents}
\label{alg:required_inputs}
\end{algorithm}

\subsection{Component Integration and Chatbot Generation}
\label{subsec:approach_componentIntegration}

The analysis and mapping phases described above have been materialized in a software tool called \textit{Demabot}. After an initial configuration this tool is in charge of generating all the elements required for generation and interacting with the external tools (sentence generation and NLU) automatically and transparently to the user. In the following section, we provide more details about its architecture. 
 \section{Demabot: A Tool for the Automatic Generation of Chatbots}
\label{sec:demabot}

\begin{figure}[thb]
  \centering
  \includegraphics[width=\linewidth]{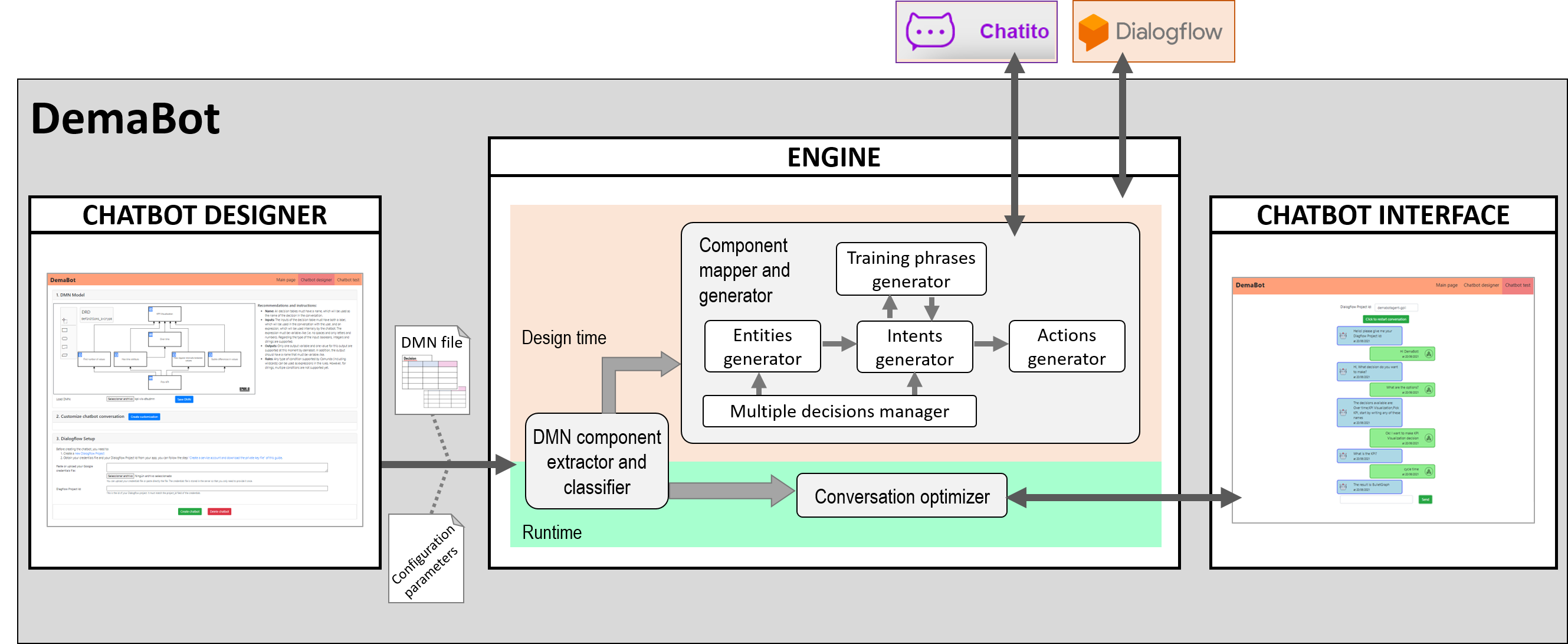}
  \caption{Demabot architecture, adapted from \cite{Estrada_2021_Demabot}} \label{fig:demabot_architecture}
\end{figure}

\textit{Demabot}\footnote{Demabot is available at: http://resisa01.us.es/} is a web tool that provides a low-code solution to create a ready-to-use \textit{decision-support chatbot} based on DMN decision models. 
Furthermore, it complies with the conversational and interaction requirements specified in Section \ref{sec:behavior}. A preliminary version of the Demabot architecture was presented in \cite{Estrada_2021_Demabot}. 
Demabot consists of three modules illustrated in Figure~\ref{fig:demabot_architecture}: \textit{chatbot designer}, \textit{engine}, and \textit{chatbot interface}.

\subsection{Demabot Chatbot Designer}
\label{subsec:demabot_designer}

\begin{figure}[thb]
  \centering
  \includegraphics[width=\linewidth]{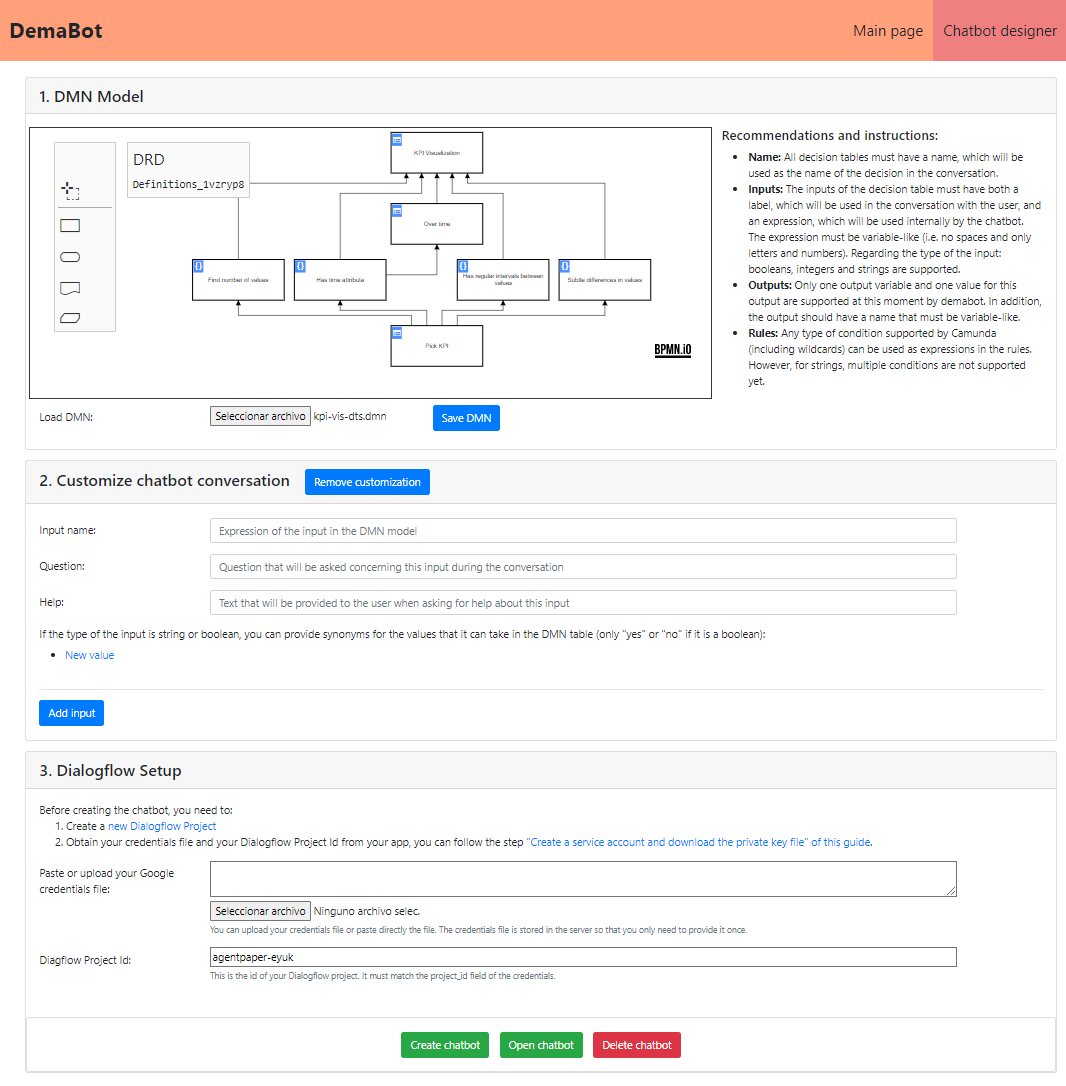}
  \caption{Demabot - Chatbot designer module} \label{fig:demabot_tool_designer}
\end{figure}

This module supports the definition of the inputs of Demabot. The inputs of Demabot are a \textit{DMN file} containing a DMN decision model, and \textit{configuration parameters} including the customization of the chatbot conversation and the \textit{credentials file} associated with a Dialogflow project where the generated chatbot components will be stored. Figure \ref{fig:demabot_tool_designer} shows the interface of this module. 

\textit{Chatbot designer} provides three functionalities. First, the DMN editor (\textit{1.~DMN model} in Fig.~\ref{fig:demabot_tool_designer}) integrated with Camunda\footnote{https://camunda.com/} from which a DMN decision model can be generated, or an existing \textit{DMN file} can be imported. Second, the possibility to customize the chatbot conversation by configuring the questions and answers concerning an input in a conversation turn. It also provides the possibility of defining synonyms for input entries of type string or boolean. (\textit{2.~Customize chatbot conversation} in Fig.~\ref{fig:demabot_tool_designer}). And third, the generation of the chatbot (\textit{3.~Dialogflow Setup} in Fig.~\ref{fig:demabot_tool_designer}), which requires loading the \textit{credentials file} and providing the \textit{project ID}.


\subsection{Demabot Engine}
\label{subsec:demabot_engine}

The engine was implemented as a web API developed in JAVA under the SpringBoot framework and Maven. In addition, the use of other libraries is required: \textit{Camunda-engine} is used to read and process the DMN decision models, \textit{Google-cloud-dialogflow} to list, create, edit and delete the elements necessary for the creation of the chatbot; and \textit{Google-api-client} to manage Google credentials, since without them it is not possible to use any of the Dialogflow functionalities. 

At design time, this module processes the inputs received from the \textit{Chatbot Designer} and interacts with Dialogflow to generate and store all chatbot components (entities, intents, etc.). \textit{Engine} also connects with Chatito\footnote{Chatito is available at: https://rodrigopivi.github.io/Chatito/} to build the phrases that will train the generated chatbot. At runtime, \textit{Engine} interacts with the \textit{Chatbot Interface} module and analyzes the inputs received from the user to optimize the conversation. \textit{Engine} has three submodules described~below.


\subsubsection{DMN Component Extractor and Classifier}
\label{subsubsec:demabot_engine_extractor}
This submodule identifies the DMN components (decisions, rules, inputs and outputs) and classifies the expected values and types for each component from the \textit{DMN file}. At design time, these data, the \textit{credentials file} and the \textit{project ID} are sent to the \textit{Component mapper and generator}. At run time, the data is sent to the \textit{Conversation optimizer}.


\subsubsection{Component Mapper and Generator}
\label{subsubsec:demabot_engine_generator}
This submodule receives the classified DMN elements and identifies the chatbot components to automatically generate for each of them following characteristics described in Sect.~\ref{subsec:approach_mappingComponents}, so it has a \textit{generator} for each component. 
\textit{Entities generator} creates an entity for each type of attribute not supported by the platform; \textit{Intents generator} creates an intent for each DMN input, as well as generating help intents to give additional information to the user, for example the list of decisions available to be made, or the types of values expected for an input; \textit{Actions generator} specifies the actions to be performed, including the responses to be sent to the user; \textit{Multiple decision manager} establishes relationships between decision tables in hierarchical structures when the output of one table is the input of another; and \textit{Training phrases generator} uses the Chatito engine to build training phrases that will help the user interact with the chatbot. These phrases are automatically generated using the parameters extracted from the \textit{DMN file} and optionally, from user-defined phrase configuration parameters.


\subsubsection{Conversation Optimizer}
\label{subsubsec:demabot_engine_optimizer}
This submodule interacts with the \textit{extractor and classifier} submodule at runtime to communicate with the Dialogflow platform to recognize the intents, parameters and entities from the user utterances. It analyzes the information provided by DialogFlow to identify the required inputs to make a decision. It also applies an optimization algorithm to ask for input values only if the input is required depending on the values captured so far, and recognizes several parameters (inputs) received in a single user message, thus minimizing the number of interactions to produce the decision.


\subsection{Demabot Chatbot Interface}
\label{subsec:demabot_interface}

\begin{figure}[thb]
  \centering
  \includegraphics[width=0.9\linewidth]{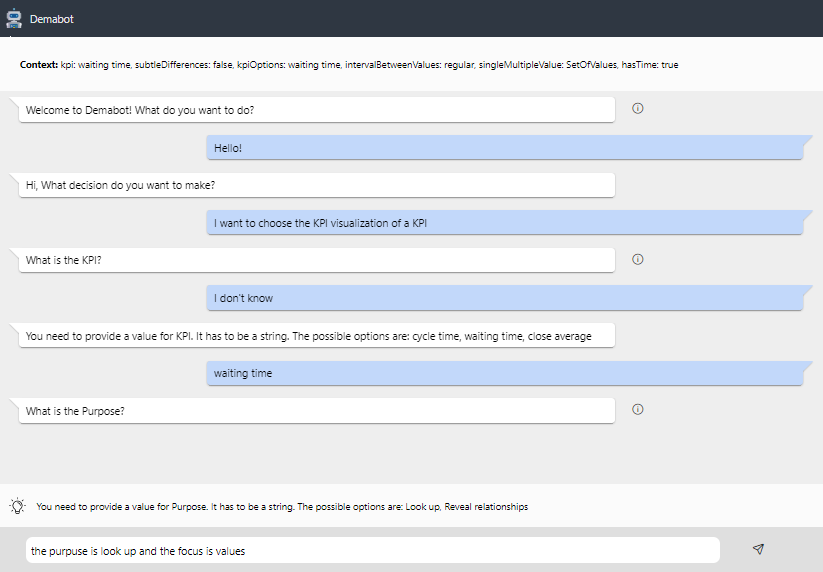}
  \caption{Demabot - Chatbot interface module} \label{fig:demabot_tool_interface}
\end{figure}

This module provides an environment for the communication between the user and the generated chatbot, which is shown in Figure~\ref{fig:demabot_tool_interface}. It communicates with the \textit{Conversation optimizer} to minimize conversation turns with the user. The chatbot interface can also be used as a chatbot test web environment to check the behavior of the chatbot while designing the model.

 \section{Evaluation}
\label{sec:evaluation_scenario}

This section describes the scenario for evaluating our proposal for automatically generating decision-support chatbots. For this purpose, we conducted a workshop intending to evaluate the three characteristics that we pursue with this proposal: that the chatbot generation should require as little prior technical knowledge as possible, that the chatbot generation process should be intuitive, simple, and understandable; and that the generated chatbot should interact correctly and fluently with the final user.

\subsection{Workshop Description}
\label{subsec:evaluation_workshop_OLD}

The workshop was conducted through a video call with each participant using the \textit{Teams} platform. First, the presenter made a brief introduction about the objectives and characteristics of the workshop. Then, details about the activities to be developed during the workshop were provided and described through Google form. 
The workshop was attended by 15 participants aged between 22 and 52 years old. All of them have computer science skills, have university work experience (from six months to more than ten years) and are currently employed at a university. 
Of these participants, six have doctoral degrees, five have master's degrees and four have bachelor's degrees. No prior knowledge was required for this workshop, as the necessary information was provided during the activity. Both the form describing the activity and the questions asked in each section, as well as the results obtained are available in a public repository\footnote{Repository available at: https://github.com/Adartse/Demabot-ExtraResources}.

To generate a chatbot using Demabot, it is necessary to set up a project in Dialogflow. As this is not considered part of the automatic chatbot generation process, Dialogflow projects were configured prior to the workshop. To avoid possible synchronization conflicts, one project was created for each participant. At the beginning of the workshop, each participant was assigned a number to access a folder with the project credentials file and its ID. 
From this moment on, the participants participants started the activity and remained connected throughout the workshop. The presenter was available throughout the workshop. Specific doubts of any participant were discussed in a private conversation with the presenter so as not to interfere with the work of the others.

The workshop consisted of nine sections. After an introductory section, a section was included to obtain demographic information from the participants, followed by a section to inquire about their prior knowledge and experience with the concepts and tools related to the workshops. 
The next section provided required foundations to developed the workshop through two videos. The first video, lasting ten minutes, described the basic concepts of DMN. The second, lasting 13 minutes, introduced the Demabot tool, described how to model a decision model in it, how to generate a chatbot, how to customize it, and how to interact with it once generated. 
In the following three sections, participants were asked to generate, modify or customize chatbots following certain indications. These sections will be detailed later. 
The last section consisted of a set of questions to gather the final opinion of the participants regarding the knowledge needed to develop this activity, the attractiveness and usefulness of the chatbots and the tools used, the development process and the results obtained.
On average, participants have taken two hours and 23 minutes to develop all parts of the workshop, with the best time being 46 minutes and the longest time being three hours. 


\subsubsection{Practical exercise 1: Chatbot based on a given DMN decision model}
\label{subsubsec:evaluation_activities_gen1}

The aim of this activity was for participants to generate a decision support chatbot with Demabot based on a DMN decision model given by the authors and to be modified by the participants as directed. 
The decision model in DMN\footnote{Decision table available at: http://resisa01.us.es/, in section \textit{Chatbot Designer}.} aims to determine whether a person is \textit{accepted}, \textit{rejected} or \textit{conditionally accepted} to participate in an association depending on three criteria: age of the applicant (age), whether currently employed or not (hired), and the size of the contribution made previously (contribution). 12 rules made up the decision table. 
Participants had to modify the decision table without altering the number of input entries and taking into account that: (i) If age is $>$55 and the candidate is not currently hired, the outcome is ``conditionally accepted". (ii) If age is $>$55 and  the candidate is hired, then  the outcome is ``accepted". The changes only affect rule number 12, where the value of an input and the result must be modified to satisfy requirement (i). In addition, a new rule should be created to satisfy requirement (ii). 
After this, participants should have been able to generate the chatbot following the indications presented in the video. 
Finally, participant were asked to interact with the chatbot. For this, two types of interaction were proposed: the first, in which the participant responds to the chatbot by providing single input values, as asked by the chatbot; and the second, the user interacts with the chatbot by providing sentences that include more than one input value.


\subsubsection{Practical exercise 2: Customize inputs and generate chatbots}
\label{subsubsec:evaluation_activities_cust2}

The objective of this practical exercise was for the participant to customize the conversation by modifying the information of at least one input of type string or boolean. Participants were asked to modify the question associated with the input, to modify the information in its help bubble and to add at least one synonym for the selected input. 
After this, the chatbot was to be re-generated and verify that the changes were applied correctly.

\subsubsection{Practical exercise 3: Chatbot based on a user-generated decision model}
\label{subsubsec:evaluation_activities_gen3}

The third practical exercise aims for each participant to create their own DMN decision model from a given scenario to generate a fully functional chatbot. 

The decision model was built from the information provided by a Washing Machine Buying Guide \footnote{Guide available at: https://bit.ly/3vv87JK}. This guide consists of eight configuration steps before providing a suggestion and some of them consist of a large number of possible values to be selected.  
The decision table had to be formed by a set of decision rules that provided as output the model of a washing machine, taking as a basis for this decision the characteristics of the washing machines (inputs) offered by the manufacturer. 
Since the buying guide provides a large number of configuration options to suggest their models, participants were instructed  (i) to omit some configuration options, so they were expected to identify five inputs for their decision table, and (ii) not to include all possible decision rules, but to include at least eight of them. 
This does not affect the purpose of the exercise, as the end result of this phase is still a decision model upon which the chatbot will be built. 
The exercise was prepared for the participant to identify string, boolean and integer inputs, however the expected input type was not indicated to test the flexibility of Demabot. 





\subsection{Results}
\label{subsec:evaluation_results}

In this section, we summarize the results of both evaluation processes.



\begin{figure}[thb]
    \centering
	\includegraphics[trim={0cm 0cm 0cm 0cm},clip,width=\linewidth]{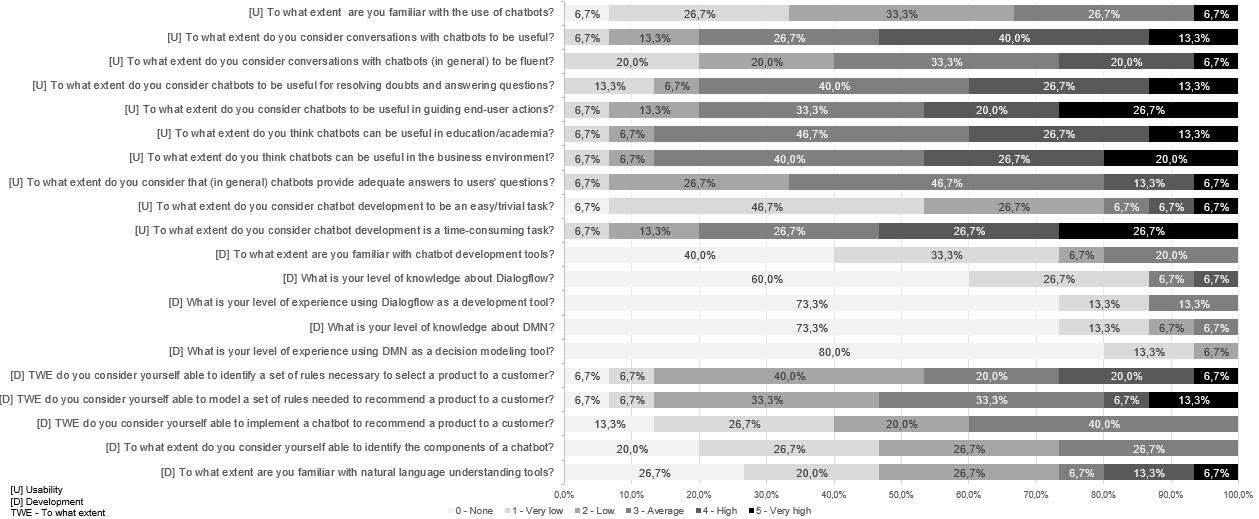}
	\caption{Previous knowledge of the participants about usability and chatbots development. }
	\label{fig:graph-sect-02}       
\end{figure}

Figure \ref{fig:graph-sect-02} shows the results related to participants' prior knowledge and experience regarding chatbot usability [U] and knowledge about Demabot-related development tools [D]. 

As aspects to highlight related to usability, only 6.7\% of the participants stated that they were very familiar with the use of chatbots, the rest reflected an average, low, very low or no knowledge. They have a very varied opinion regarding the usefulness and fluency of conversations with chatbots. Most of them consider that chatbots can be useful both in academia and in business environments. About 80\% consider that chatbots development is not an easy task and that on the contrary it is a time-consuming task. With respect to development, 100\% of the participants stated that they had no (40\%), low (33\%), very low (6,7\%) or average (20\%) knowledge of chatbot development tools. Only 6.7\% indicated having a high knowledge of Dialogflow, and 100\% indicated having experience in the use of Dialogflow between null and average. Something similar happens with the DMN standard, where 100\% of participants indicated having average or below average knowledge, and between low and null with respect to experience in the use of the standard. Similarly, more than 70\% indicate having low or below average knowledge about the use of NLU tools.

\begin{figure}[thb]
    \centering
	\includegraphics[trim={0cm 0cm 0cm 0cm},clip,width=\linewidth]{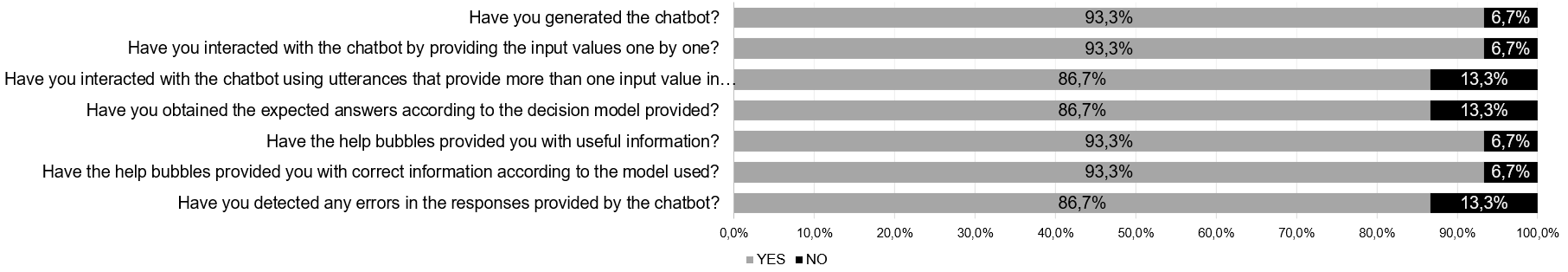}
	\caption{Results of the first practical exercise of chatbot generation using Demabot.}
	\label{fig:graph-sect-04}       
\end{figure}

Regarding the \textit{first practical exercise}, 93,3\% of the participants were able to generate the chatbot and interact with it by providing the values one by one, as requested during the conversation. In addition, all of them considered that the help bubbles provided useful information in line with the decision model used. Only one of the participants reported having problems during the interaction with the chatbot when providing more than one input at a time. Although it was not specified on the response form, the participant who was unable to generate the chatbot after the decision rules were modified incurred a modeling error that was not identified until the end of the exercise. The complete set of questions and its answers are collected in  Figure~\ref{fig:graph-sect-04}. 

As for the \textit{second practical exercise}, out of the 15 participants, 14 performed the sugested customization. Of these, 50\% customized the input \textit{contribution} of type string, while 50\% did it on the input hired of type boolean. 
Of participants who applied the customization, 100\% indicated both the customization of the question and the information of the help bubble were performed correctly. Regarding the definition of the synonyms, 21\% of participants (three) indicated that they had problems on this point. Specifically that the synonym used was not recognized by the chatbot. In one of these error cases, the input contribution was customized by using values such as \textit{0} and \textit{1} as synonyms for the values \textit{none}, \textit{minimum}, etc. This apparently created a conflict in the parsing of the sentences because of the data types used. In the other two cases, the root cause of the problem was not identified on the input hired, as indicated by the other two participants. 

The \textit{third exercise} had two parts, the first is the construction of the decision model and the second, the configuration and generation of the chatbot. 
Although it was indicated that about eight rules should be modeled, on average 11 rules were identified and ten were modeled. However, although most participants modeled eight rules, the minimum was two and the maximum was 23. 
Out of the 15 participants, 53,3\% identified five inputs, 26,7\% identified four inputs and the remaining 20\%, between seven and eight inputs. The difference may be due to confusion between the steps required by the buying guide (eight) and the specific entries requested during the exercise (five).

With respect to modeling, again 53,3\% modeled a decision table with five inputs and 26,7\% with four inputs, 6,7\% modeled two inputs, and 13,3\% modeled seven inputs. The differences between identified and modeled inputs derive from the fact that some participants identified that certain input values were not relevant to obtain the decision output and omitted them from the final model. Participants who modeled seven inputs included inputs that were suggested to be omitted. 
There was no single way to model the decision table. Out of the five expected inputs (\textit{Household people}, \textit{fabrics}, \textit{noise level}, \textit{efficiency}, and \textit{programs}) only\textit{ Household people} was specified to be a numeric value, since each input entry would contain a range of numeric values (e.g., between one and two persons, [1..2]). The rest of the entries could be defined as a string or boolean.
Three types of inputs (integer, string, boolean) were used by 53.3\% of the participants, 33.3\% used only two of these types (integer and boolean or string and boolean) and 13.3\% defined all values as string. This variation in the way of defining the inputs of the decision table does not affect the chatbot generation process, it only modifies the values allowed when interacting with it. The decision models created by the participants are also available in the online repository. 

Finally, a questionnaire was conducted to find out the users' opinion regarding the prior knowledge needed to perform the activity, the attractiveness/usefulness of the tool, the development of a decision-support chatbot using Demabot, about the overall performance of the chatbot, as well as some open questions to express the general opinion about the tool and the chatbot generation process. 

\begin{figure}[thb]
    \centering
	\includegraphics[trim={0cm 0cm 0cm 0cm},clip,width=\linewidth]{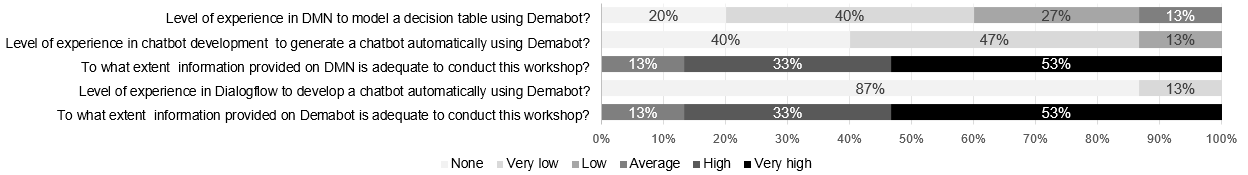}
	\caption{Final results related to the background knowledge required for the use of Demabot. }
	\label{fig:graph-sect-08A}       
\end{figure}

Regarding the background (Fig.~\ref{fig:graph-sect-08A}), the majority indicates that the previous knowledge about chatbot development required to use Demabot is very low (47\%), none (40\%) or low (13\%); while about DMN 40\% also indicate that it is very low, low (27\%) or none (20\%); and about the knowledge about Dialogflow 87\% indicate that the required level is none and 13\% that it is very low. The majority agrees that the information provided is adequate to develop the workshop. 

\begin{figure}[thb]
    \centering
	\includegraphics[trim={0cm 0cm 0cm 0cm},clip,width=\linewidth]{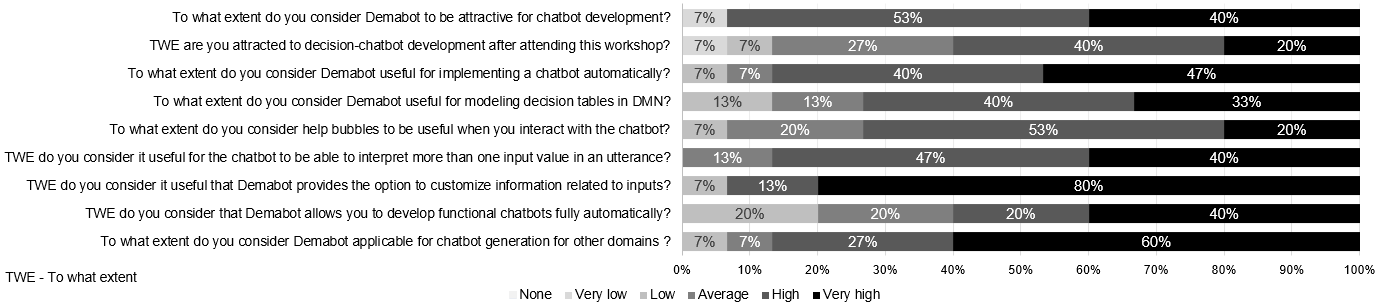}
	\caption{Final results related to the attractiveness and usefulness in use of Demabot. }
	\label{fig:graph-sect-08B}       
\end{figure}

From the answers presented in Fig.~\ref{fig:graph-sect-08B}  we can extract that after the workshop the participants consider that both the development of chatbots and the Demabot tool are attractive to them. In particular, they consider Demabot to be really useful for the automatic implementation of chatbots (47\%very high and 40\% high) and for decision modeling (40\% high and 33\% very high). Moreover, additional features such as help bubbles (53\% high and 20\% very high), recognition of multiple inputs in an utterance (47\% high and 40\% very high) and input customization (80\% very high and 13\% high) are also considered useful. In addition, most of them consider that the tool can be used in a different environment (60\% very high and 27\% high).

\begin{figure}[thb]
    \centering
	\includegraphics[trim={0cm 0cm 0cm 0cm},clip,width=\linewidth]{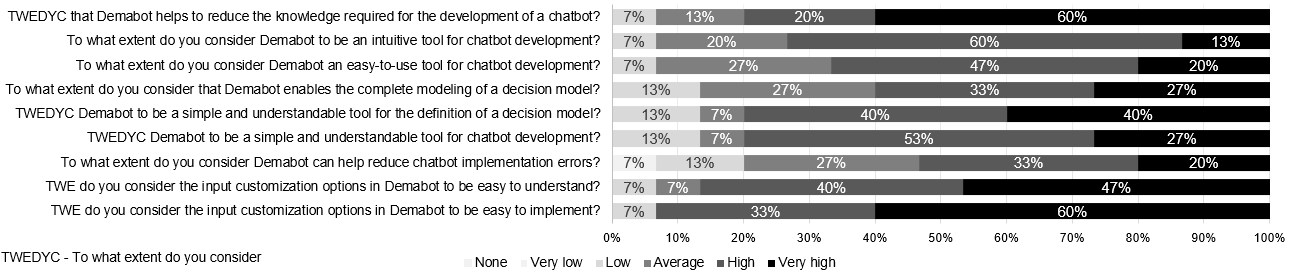}
	\caption{Final results related to the chatbot development using Demabot. }
	\label{fig:graph-sect-08C}       
\end{figure}

Regarding development (Fig.~\ref{fig:graph-sect-08C}), 80\% of the participants agree (60\%very high and 20\% high) that Demabot helps to reduce the knowledge needed to develop chatbots. They also consider it an intuitive tool (60\% high), easy to use (47\% high and 20\% very high) and that it is simple and understandable (53\% high and 27\% high). In particular, the customization of inputs is considered easy to understand (47\% very high) and easy to implement (60\%). 

Translated with www.DeepL.com/Translator (free version)

\begin{figure}[thb]
    \centering
	\includegraphics[trim={0cm 0cm 0cm 0cm},clip,width=\linewidth]{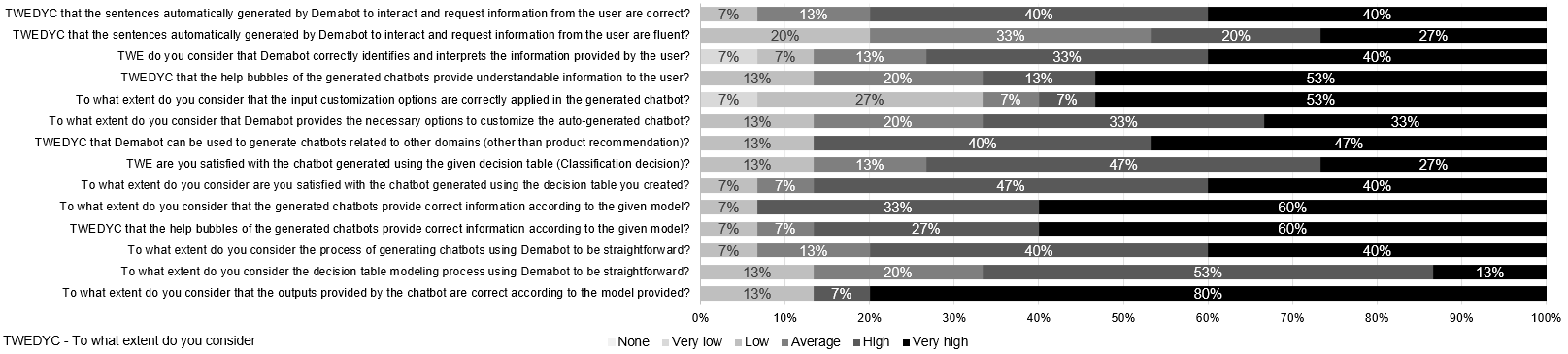}
	\caption{Final comments on the generated chatbots and the automatic generation process. }
	\label{fig:graph-sect-08D}       
\end{figure}

Finally, from the results shown in Fig.~\ref{fig:graph-sect-08D} we can extract that the participants consider that it is possible to interact with Demabot correctly and fluently, and that it is able to interpret the information provided by the user. That its customization and help features are understandable and correct. The majority of them are satisfied with the chatbots generated, both the one based on a given model (47\% high and 27\% very high), and the one generated with the model generated by them (47\% high and 40\% very high). They are much more satisfied with the one generated by themselves. They strongly believe that the chatbot provides correct information according to the given model (60\% very high) and they consider the chatbot generation process to be straightforward (40\% very high and 40\% high). 

 \section{Related Work}
\label{sec:related_work}

Chatbots are currently a trending topic, so there are a large number of articles describing its use in various areas. In the context of this article, we focus exclusively on proposals that describe methods for generating chatbots.

The chatbots is a concept closely related to the concept of robotic process automation (RPA), which refers to software agents acting as humans in system interactions, and process automation. 
Syed et al.~\cite{Syed_2020_RPA} in their analysis about challenges related to RPA highlighted the lack of methodological support for their implementation, as well as the systematic design, development and evolution. 
Our proposal addresses the first three challenges by providing a systematic solution for developing fullyfunctional chatbots automatically, which reduces errors, time and effort in the development process. 

Regarding the generation of chatbots, Chitto et al. \cite{Chitto_2020_HTMLChatbots} presented a preliminary work describing a set of abstractions, techniques and conceptual vocabulary for the semi-automatic generation of chatbots from information obtained from the web sites themselves. 
In \cite{Banisharif_2022_BusinessChatbots}, a framework is proposed for the automatic generation of business chatbots on the basis of a domain-specific language and templates based on JSON language. Although it is noted that the chatbot generated is built based on a proposed metamodel, no information is provided on the level of detail of the ground data on which the chatbot is built, nor on the fluency or customization of the generated result. 
Gwendal et al. \cite{Gwendal_2019_Framework} highlighted the need for in-depth advanced technical knowledge for the development and deployment of chatbots, which increases costs. 
They proposed a multi-platform chatbot modeling framework and provided a domain specific language to define chatbots in a platform-independent way. 
During this chatbot generation process, a developer must specify the chatbot components through a modeling language where intents, actions, and the interaction between them are defined. 
Therefore, this process still requires a high level of technical knowledge on the part of the user-developer. 

In the context of business processes, L\'opez et al. \cite{Lopez_2019_ProcModelChatbot} described a methodology and a prototype to transform a BPMN model into a chatbot based on AIML language\footnote{\url{http://www.aiml.foundation/}}. Although its evaluation revealed shortcomings in the fluency and certainty of the chatbot responses, it also shows the high potential of the proposal and the multiple possibilities of extension.

In the particular case of chatbot generation from decision models, the proposal in \cite{Etikala_DMNChats_2022} describes a set of scenarios during the conversation between the chatbot and a user, but it is not clear the level of automation achieved, since it is intuited the need for the participation of a user as developer. 
Our proposal improves on the previous one in that we propose a fully automatic transformation process where user participation is required only during the modeling or definition of the decision model. Our proposal also includes the integration with external tools such as NLU platforms and the automatic generation of training sentences in a completely transparent way for the user. In addition to providing the help features that are not completely covered in other proposals.

 \section{Conclusions and Future Work}
\label{sec:conclusions_futurework}

In this article, we present a novel approach to automatically generate chatbots using decision models based on DMN. The approach includes a methodology for identifying and mapping elements of a DMN model for use within a chatbot agent context and a software tool, Demabot, that implements that methodology. The chatbot generation process considers a set of conversational requirements defined based on characteristics associated with conversational behavior that should be present in chatbots (conscientiousness, proactivity, and communicability) that were identified in the literature.

The feasibility and correctness of our approach were evaluated through a workshop that included the participation of 15 people.
We conclude that the proposal is feasible, is domain-independent, and can be used by users with little technical knowledge of both DMN and chatbot development tools. 
Participants found the process of chatbot generation easy to follow, and the results obtained were deemed correct according to the defined and provided requirements.
While the results were considered correct, it is worth noting that the success in the accuracy of the generated chatbot largely depends on the quality of the DMN model and the decision rules from which it is built.
Future work aims to adapt the transformation methodology and the generated tool to integrate more powerful text analysis tools. These changes mainly concern the construction of the training phrases required for the chatbot to achieve fluid communication.

\bibliography{Demabot}

\begin{thebibliography}{10}
\expandafter\ifx\csname url\endcsname\relax
  \def\url#1{\texttt{#1}}\fi
\expandafter\ifx\csname urlprefix\endcsname\relax\def\urlprefix{URL }\fi
\expandafter\ifx\csname href\endcsname\relax
  \def\href#1#2{#2} \def\path#1{#1}\fi

\bibitem{Hussain_2019_SurveyChatbots}
S.~Hussain, O.~Ameri~Sianaki, N.~Ababneh, {A Survey on Conversational Agents}, in: Proceedings of {WAINA}, 2019, pp. 946--956.

\bibitem{Valtolina_2108_DomainOfUse}
S.~Valtolina, B.~R. Barricelli, S.~D. Gaetano, P.~Diliberto, {Chatbots and Conversational Interfaces}, in: Proceedings of {CoPDA}, 2018, pp. 62--70.

\bibitem{Janssen_2022_Chatbots}
A.~Janssen, D.~R. Cardona, J.~Passlick, M.~H. Breitner, How to make chatbots productive, Int. J. Hum. Comput. Stud. 168 (2022) 102921.

\bibitem{Sorin_2021_RepetitiveDecs}
S.~Anagnoste, I.~Biclesanu, F.~D'Ascenzo, M.~Savastano, The role of chatbots in end-to-end intelligent automation and future employment dynamics, in: Business Revolution in a Digital Era, 2021, pp. 287--302.

\bibitem{Seeber_2020_chatbotsMates}
I.~Seeber, E.~Bittner, et~al., {Machines as teammates: A research agenda on AI in team collaboration}, Inf. Manag. 57~(2) (2020) 103174.

\bibitem{Jovanovic_2021_ChatbotsHealth}
M.~Jovanovic, M.~B{\'{a}}ez, F.~Casati, Chatbots as conversational healthcare services, {IEEE} Internet Comput. 25~(3) (2021) 44--51.

\bibitem{Mujeeb_2017_Aaquabot}
S.~Mujeeb, M.~H. Javed, T.~Arshad, Aquabot: a diagnostic chatbot for achluophobia and autism, Int. J. Adv. Comput. Sci. Appl. 8.

\bibitem{web_chatbot_vasco}
G.~V. Depto.~de Salud, {Osakidetza - Protocolo COVID-19}, \url{https://www.osakidetza.euskadi.eus/}, accessed: 07-07-2020 (2020).

\bibitem{web_chatbot_riskassessment}
I.~LiveChat, {COVID-19 Risk Assessment Chatbot}, \url{https://www.chatbot.com/covid19-chatbot/}, accessed: 15-10-2022 (2022).

\bibitem{web_chatbot_hispabot}
M.~d. A. E. y. T.~D. Gobierno~de España, {HISPABOT-COVID19}, \url{https://covid19.gob.es/hispabot-covid19}, accessed: 15-10-2022 (2022).

\bibitem{Brandtzaeg_2017_WhyUseChatbot}
P.~B. Brandtzaeg, A.~F{\o}lstad, {Why People Use Chatbots}, in: Internet Science, 2017, pp. 377--392.

\bibitem{Shawar_2007_Chatbotuseful}
B.~A. Shawar, E.~Atwell, {Chatbots: Are they really useful?}, in: Proceedings of {LDV} {Forum}, Vol.~22, 2007, pp. 29--49.

\bibitem{Fish_2012_AutomationKnowledge}
A.~Fish, Knowledge Automation: How to Implement Decision Management in Business Processes, 1st Edition, Wiley, United States, 2012.

\bibitem{OMG_DMN_2019}
{Object Management Group}, {Decision Model and Notation (DMN) specification. V1.3}, \url{https://www.omg.org/spec/DMN}, {Access: 20-08-2022} (2019).

\bibitem{Figl_2018_DMN}
K.~Figl, J.~Mendling, G.~Tokdemir, J.~Vanthienen, What we know and what we do not know about {DMN}, Enterp. Model. Inf. Syst. Archit. Int. J. Concept. Model. 13 (2018) 2:1--16.

\bibitem{Dumas_2018_Fundamentals}
M.~Dumas, M.~La~Rosa, J.~Mendling, H.~A. Reijers, Fundamentals of Business Process Management, 2nd Edition, Springer, Berlin, Heidelberg, 2018.

\bibitem{Ahmad_2020_Chatbots}
A.~Abdellatif, D.~Costa, K.~Badran, R.~Abdalkareem, E.~Shihab, {Challenges in Chatbot Development}: {A} study of stack overflow posts, in: Proceedings of {MSR} '20, 2020, pp. 174--185.

\bibitem{Gwendal_2019_Framework}
G.~Daniel, J.~Cabot, L.~Deruelle, M.~Derras, {Multi-platform Chatbot Modeling and Deployment with the Jarvis Framework}, in: Proceedings of CAiSE, 2019, pp. 177--193.

\bibitem{Unal_2019_KPI}
{\"{U}}.~Aksu, A.~del{-}R{\'{\i}}o{-}Ortega, M.~Resinas, H.~A. Reijers, An approach for the automated generation of engaging dashboards, in: {OTM} 2019 Conferences, 2019, pp. 363--384.

\bibitem{Chavez_2021_Challenges}
A.~P. Chaves, M.~A. Gerosa, How should my chatbot interact? a survey on social characteristics in human–chatbot interaction design, Int. J. Hum. Comput. Interact. 37~(8) (2021) 729--758.

\bibitem{Estrada_2021_Demabot}
B.~Estrada{-}Torres, A.~del{-}R{\'{\i}}o{-}Ortega, M.~Resinas, Demabot: a tool to automatically generate decision-support chatbots, in: {BPM} 2021, Vol. 2973 of {CEUR} Workshop Proceedings, 2021, pp. 141--145.

\bibitem{Syed_2020_RPA}
R.~Syed, S.~Suriadi, M.~A. et~al., Robotic process automation: Contemporary themes and challenges, Computers in Industry 115 (2020) 103162.

\bibitem{Chitto_2020_HTMLChatbots}
P.~Chitt{\`o}, M.~Baez, F.~Daniel, B.~Benatallah, Automatic generation of chatbots for conversational web browsing, in: Conceptual Modeling, Springer International Publishing, Cham, 2020, pp. 239--249.

\bibitem{Banisharif_2022_BusinessChatbots}
M.~Banisharif, A.~Mazloumzadeh, M.~Sharbaf, B.~Zamani, {Automatic Generation of Business Intelligence Chatbot for Organizations}, in: Proceedings of {CSICC} 2022, {IEEE}, 2022, pp. 1--5.

\bibitem{Lopez_2019_ProcModelChatbot}
A.~L{\'o}pez, J.~S{\`a}nchez-Ferreres, J.~Carmona, L.~Padr{\'o}, {From Process Models to Chatbots}, in: Proceedings of CAiSE, 2019, pp. 383--398.

\bibitem{Etikala_DMNChats_2022}
V.~Etikala, A.~Goossens, Z.~Van~Veldhoven, J.~Vanthienen, Automatic generation of intelligent chatbots from dmn decision models, in: Rules and Reasoning, Springer International Publishing, Cham, 2021, pp. 142--157.

\end{thebibliography}

\end{document}